%% file: 0_main.tex
\documentclass{Interspeech}

\usepackage{pifont}
\usepackage{multirow}
\usepackage{microtype}
\usepackage{graphicx}
\usepackage{acronym}
\usepackage{subfigure}
\usepackage{booktabs}
\usepackage{hyperref}

\usepackage{colortbl}
\usepackage{soul}
\usepackage[dvipsnames]{xcolor}
\usepackage{color, xcolor} 
\usepackage[utf8]{inputenc}
\usepackage{algorithm}
\usepackage{algpseudocode}
\usepackage{amsmath}
\usepackage{amssymb}
\usepackage{mathtools}
\usepackage{amsthm}
\usepackage[capitalize,noabbrev]{cleveref}
\theoremstyle{plain}

\theoremstyle{definition}

\theoremstyle{remark}

\usepackage[textsize=tiny]{todonotes}

\interspeechcameraready 

\title{Efficient and Microphone-Fault-Tolerant 3D Sound Source Localization}

\author[]{Yiyuan}{Yang}
\author[]{Shitong}{Xu}
\author[]{Niki}{Trigoni}
\author[]{Andrew}{Markham}

\affiliation[nocounter]{Department of Computer Science}{University of Oxford}{United Kingdom}
\email{\{yiyuan.yang,shitong.xu,niki.trigoni,andrew.markham\}@cs.ox.ac.uk}

\usepackage{comment}

\begin{document}

\maketitle

\begin{abstract}
Sound source localization (SSL) is a critical technology for determining the position of sound sources in complex environments. However, existing methods face challenges such as high computational costs and precise calibration requirements, limiting their deployment in dynamic or resource-constrained environments. This paper introduces a novel 3D SSL framework, which uses sparse cross-attention, pretraining, and adaptive signal coherence metrics, to achieve accurate and computationally efficient localization with fewer input microphones. The framework also supports operational microphones at unknown positions: their recordings remain available and are used to estimate both source and microphone positions. Preliminary experiments demonstrate its scalability for multi-source localization without requiring additional hardware. This work advances SSL by balancing the model's performance and efficiency and improving its robustness for real-world scenarios.
\end{abstract}

\input{1_introduction}
\input{2_related_work}
\input{3_method}
\input{4_experiment}
\input{5_conclusion}

\clearpage
\bibliography{6_reference}
\bibliographystyle{IEEEtran}

\end{document}

%% file: 1_introduction.tex
\section{Introduction} \label{sec:introduction}

Sound source localization (SSL) involves accurately determining the direction and distance of the sound sources in complex and noisy environments~\cite{popper2005sound}. It is crucial for enhancing situational awareness, facilitating human-computer interaction, and optimizing the performance of various signal-processing applications. For instance, prior works have incorporated sound source localization modules to improve speech quality in online meetings and virtual reality, improve speech intelligibility in hearing aids, and identify noise sources in surveillance and acoustic imaging~\cite{desai2022review}.

In the early stages, due to hardware limitations and experimental constraints, traditional sound source localization was primarily conducted in two-dimensional space. However, 2D localization only determines the azimuth angle within the horizontal plane, and neglects useful information such as elevation and distance of the sound source~\cite{grumiaux2022survey}. Since we live in a three-dimensional world, the advancement of technology and increasing application demands have driven the exploration toward 3D sound source localization, which learns a more comprehensive and precise spatial representation of the 3D scene. 

Existing 3D SSL algorithms typically rely on densely deployed microphone arrays to estimate the direction of arrival (DOA) using time-difference-of-arrival (TDOA) techniques or beamforming algorithms as the input features~\cite{yalta2017sound}. While these methods perform well in controlled environments, they require precise microphone calibration, which incurs high computational costs and imposes strict sensor requirements to prevent catastrophic localization failures. These constraints significantly hinder the deployment of SSL in resource-constrained or dynamic environments, particularly when microphone availability and reliability cannot be guaranteed~\cite{gong2022end,lee2021deep}.

In practical applications, 3D sound source localization faces several critical challenges that hinder its effectiveness and scalability. Firstly, computational efficiency is paramount, as real-time processing and low inference latency are essential for dynamic environments such as live communication platforms~\cite{kapoor2016novel}. Secondly, reducing the number of microphones required for accurate localization is crucial to lower hardware costs and simplify deployment, especially in resource-constrained settings~\cite{munirathinam2024sound}. In addition, the coordinates of some operational microphones may be unavailable, which may degrade localization accuracy~\cite{gong2022end}. Finally, the ability to simultaneously localize multiple overlapping sound sources is a significant challenge, as it requires sophisticated algorithms to disentangle and process concurrent audio signals without increasing hardware complexity. Addressing these issues is vital for advancing SSL technologies toward practical, real-world applications.

\input{figure/task}

As shown in Figure~\ref{fig:task}, this paper presents a novel 3D spatial sound source localization framework designed to overcome four key limitations of existing systems. (1) By leveraging sparse cross-attention and pretraining techniques, our approach significantly enhances computational efficiency and inference speed without compromising accuracy. (2) The proposed method integrates adaptive signal coherence metrics (ASCM), enabling the model to achieve the same localization precision as the traditional methods while using fewer microphones. (3) The framework uses recordings from operational microphones at unknown positions to jointly localize the source and those microphones. (4) Preliminary experiments demonstrate the framework’s scalability in handling overlapping sound events, highlighting its potential for simultaneous multi-source localization without introducing additional hardware complexity.

%% file: figure/task.tex
\begin{figure}[!t]
\centering
\includegraphics[width=0.99\linewidth]{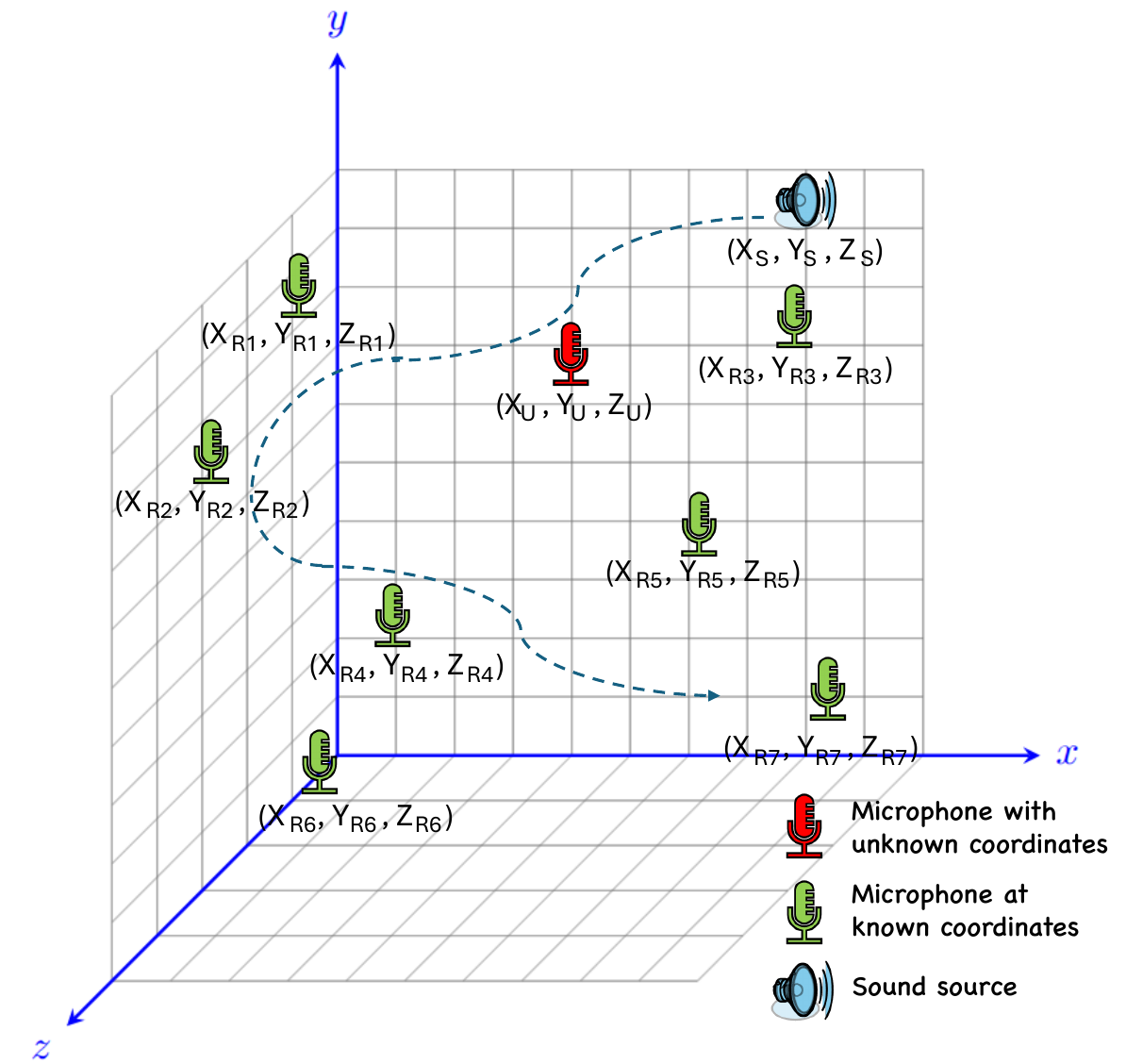} 
\vspace{-1mm}
\caption{Overview of the sound source localization task. It predicts the sound source position $[X_S, Y_S, Z_S]$ from recordings captured by calibrated microphones at known positions $[X_{Ri}, Y_{Ri}, Z_{Ri}]$. For operational microphones whose positions are unknown, the model also estimates their positions $[X_U, Y_U, Z_U]$ from their recordings.}
\vspace{-5mm}
\label{fig:task} 
\end{figure}

%% file: 2_related_work.tex
\section{Related Work} \label{sec:Related_work}

\subsection{Traditional Sound Source Localization}
Traditional approaches to sound source localization are based on establishing geometric and statistical frameworks. In the simplest scenario, when the number of independent distance measurements equals the spatial dimension, trilateration provides an elegant closed-form solution~\cite{thomas2005revisiting,manolakis1996efficient}. However, practical implementations often yield an over-determined system, wherein the maximum likelihood formulation is adopted under the assumption of additive Gaussian noise. This formulation, while theoretically appealing, results in a nonlinear, non-smooth, and non-convex optimization problem. Consequently, numerous iterative algorithms have been proposed, each offering various convergence guarantees, alongside methods that relax the original problem by minimizing errors in squared distance measurements to obtain a linear approximation~\cite{luke2017simple, jyothi2020solvit, sirola2010closed,larsson2019optimal}.

In the context of multilateralism, the extraction of TDOA estimates is critical. The generalized cross-correlation with phase transform (GCC-PHAT) has emerged as a robust and widely used technique for TDOA estimation~\cite{knapp1976generalized}. Subsequently, these measurements are incorporated into linear or minimal solvers, typically via least squares formulations, to derive the sound source location. Despite the maturity of these classical methods, challenges persist in handling real-world conditions, such as reverberation and non-line-of-sight propagation.

\subsection{Learning-based 3D Sound Source Localization}

Recent advances in machine learning have catalyzed significant progress in sound source localization tasks, particularly in extending traditional 2D techniques to 3D settings. Early studies in this domain primarily focused on estimating TDOA or DOA using convolutional neural networks to directly regress source coordinates~\cite{salvati2021time,raina2023syncnet,9268154,cho2023sr}. Although such approaches have demonstrated promising results, their applicability is often restricted to fixed microphone array configurations, limiting flexibility in dynamic or ad-hoc environments~\cite{wang2023fn,phokhinanan2023binaural,he2022sounddoa}.

To address these limitations, more sophisticated architectures have been introduced. For instance, dual-input neural networks that simultaneously process audio signals and spatial coordinates have shown improved localization performance in controlled two-dimensional scenarios~\cite{grinstein2023dual}. Similarly, graph neural networks have been employed to accommodate variable numbers of microphones~\cite{grinstein2023graph}. It aggregates spatial features derived from GCC-PHAT representations. Besides, wav2pos has designed a masked autoencoder architecture to improve the performance of the SSL task~\cite{berg2024wav2pos}. Despite these advancements, scaling learning-based methods to fully three-dimensional environments introduces additional complexities, notably due to the exponential increase in spatial discretization and the variability of acoustic conditions in realistic settings.

A promising direction lies in the integration of classical geometric insights with contemporary deep-learning frameworks. By leveraging the precise mathematical formulations of traditional methods alongside the adaptive capabilities of neural networks, hybrid approaches may offer enhanced scalability and robustness, thereby addressing the intrinsic challenges of three-dimensional sound source localization~\cite{grumiaux2022survey}.

%% file: 3_method.tex
\section{Methodology} \label{sec:Method}
\input{figure/framework}

\subsection{Problem Definition} 
Consider a 3D acoustic environment with $M$ microphones at known coordinates $\mathbf{r}_m \in \mathbb{R}^3$, $m = 1, \dots, M$ and $K$ simultaneous sound sources located at unknown positions $\mathbf{r}_k \in \mathbb{R}^3$, $k = 1, \dots, K$ ($K \ll M$). During a short time frame of $N$ samples, each source emits a distinct signal $\mathbf{s}_k \in \mathbb{R}^N$, while every microphone records a superimposed mixture of time-delayed, convolved, and noisy observations:
\begin{equation}
\small
    \mathbf{s}_m[n] = \sum_{k=1}^{K} (\mathbf{h}_{mk} * \mathbf{s}_k)[n] + \mathbf{w}_m[n], \quad n = 1, \dots, N,
\end{equation}

\noindent where $\mathbf{h}_{mk}$ denotes the room impulse response from source $k$ to microphone $m$, and $\mathbf{w}_m$ represents additive i.i.d. Gaussian noise. The task of multiple sound source localization involves estimating the positions $\{\mathbf{r}_k\}_{k=1}^{K}$ from the multichannel recordings $\{\mathbf{s}_m\}_{m=1}^{M}$ and prior knowledge of microphone geometry $\{\mathbf{r}_m\}_{m=1}^{M}$.  

In addition to the multiple SSL task, we also perform microphone position estimation from recorded audio. Alongside the $M$ microphones with known coordinates, there are $U$ operational microphones at unknown positions $\mathbf{r}_u \in \mathbb{R}^3, u = 1, \dots, U$. Their recordings $\{\mathbf{s}_u\}_{u=1}^{U}$ remain available. The extended task jointly estimates the sound source positions $\mathbf{r}_k$ and the unknown microphone positions $\mathbf{r}_u$ from the known microphone positions $\mathbf{r}_m$ and recordings at both known and unknown microphone positions. We tune our model using semi-supervised training and report its performance in Section~\ref{sec:abtest}.

\subsection{Proposed Framework}
In this subsection, we will introduce the proposed end-to-end framework, as illustrated in Figure~\ref{fig:framework}. To provide a clearer explanation, we divide it into three parts based on whether the module is designed to process the audio information only (\textcolor{LimeGreen}{Acoustic stream}), the spatial coordinate information only (\textcolor{Thistle}{Coordinate stream}), or information fusion from the audio and coordinate stream (\textcolor{YellowOrange}{Joint stream}). 

\subsubsection{Acoustic Stream} \label{AcousticStream}
In the acoustic stream, shown in Figure~\ref{fig:framework} \textcolor{LimeGreen}{green parts}, we encode each microphone recording with a pretrained audio encoder. In detail, we selected the pretrained model BEATs~\cite{chen2022beats}, which efficiently transforms raw audio into audio embedding $\mathbf{s}^{emb}_m = \mathsf{BEATs}(\mathbf{s}_m)$. Owing to its comprehensive pretrained framework, it does not require additional training, thus ensuring high data efficiency. 

\subsubsection{Coordinate Stream} \label{CoordinateStream}
For the Coordinate Stream, shown in Figure~\ref{fig:framework} \textcolor{Thistle}{pink parts}, our goal is to integrate the sound received by the microphones with their spatial coordinates to enable sound source localization.

First, for the original microphone spatial coordinates $\{\mathbf{r}_m\}_{m=1}^{M}$, we apply a simple position encoder (a two-layer MLP) to project them into a higher-dimensional position embedding $\mathbf{p}^{emb}_m$. On the other hand, we input the raw sound signals $\{\mathbf{s}_m\}_{m=1}^{M}$ into the NGCC-PHAT~\cite{berg2022extending} to extract the TDOA features. Given recordings $\mathbf{s}_i(f)$ and $\mathbf{s}_j(f)$ in the frequency domain, the generalized cross-correlation with phase transform is computed as:

\begin{equation}
\small
\mathbf{r}^{\mathrm{PHAT}}_{ij}(\tau) = \mathcal{F}^{-1} \left\{ \frac{\mathbf{s}_i(f) \mathbf{s}_j^*(f)}{|\mathbf{s}_i(f) \mathbf{s}_j^*(f)|} \right\}.
\end{equation}

\noindent The NGCC-PHAT model further refines it by applying $P$ learned filters, yielding a transformed correlation function $\hat{\mathbf{r}}^{\mathrm{PHAT}}_{ij}(\tau) = \sum_{p=1}^{P} w_p \mathbf{r}_{ij}^{\mathrm{PHAT}(p)}(\tau)$. Instead of selecting the peak value, we use the full correlation response as the TDOA feature vector $\mathbf{f}_{ij} = [\hat{\mathbf{r}}_{ij}^{\mathrm{PHAT}}(\tau_1), \hat{\mathbf{r}}_{ij}^{\mathrm{PHAT}}(\tau_2), \dots, \hat{\mathbf{r}}_{ij}^{\mathrm{PHAT}}(\tau_L)]$, where $L$ indicates the discrete time step.

To enhance the reliability of this feature, we introduce adaptive signal coherence metrics (ASCM), where the coherence with the power spectral density $g$ between microphone pairs is computed as:

\begin{equation}
\small
c_{ij}(f) = \frac{|g_{ij}(f)|^2}{g_{ii}(f) g_{jj}(f)}.
\end{equation}

\noindent This coherence measure is used to weight different microphone pairs and get the updated weighted TDOA features:

\begin{equation}
\small
\mathbf{f^{w}} = \sum_{i,j} c_{ij}^{\alpha} \mathbf{f}_{ij}.
\end{equation}

Finally, after a simple audio TDOA encoder (one-layer MLP) with the weighted TDOA features $\mathbf{f^{w}}$ as the input, we get the final audio TDOA embedding $\mathbf{r}^{emb}_{ij}$.

\subsubsection{Joint Stream} \label{JointStream}
For the joint stream, as shown in Figure~\ref{fig:framework} \textcolor{YellowOrange}{orange parts}, we integrate the three above-processed embeddings (i.e., audio embedding $\mathbf{s}^{emb}$, position embedding $\mathbf{p}^{emb}$, and audio TDOA embedding $\mathbf{r}^{emb}$) and apply various fusion strategies to facilitate sound source reconstruction and achieve our primary sound source localization task.

To effectively fuse the heterogeneous embeddings, we first employ cross-attention between the audio embeddings $\mathbf{s}^{emb}_m$ and position embeddings $\mathbf{p}^{emb}_m$. Let $\mathbf{q}$, $\mathbf{k}$, and $\mathbf{v}$ denote the query, key, and value, respectively. For the audio-to-position fusion, we derive $\mathbf{q}$ from $\mathbf{s}^{emb}_m$ and $\mathbf{k}$, $\mathbf{v}$ from $\mathbf{p}^{emb}_m$:

\begin{equation}
\small
\mathbf{q} = \mathbf{w}_q \mathbf{s}^{emb}_m, \quad \mathbf{k} = \mathbf{w}_k \mathbf{p}^{emb}_m, \quad \mathbf{v} = \mathbf{w}_v \mathbf{p}^{emb}_m,
\end{equation}

\noindent where $\mathbf{w}_q$, $\mathbf{w}_k$, and $\mathbf{w}_v$ are learnable projection matrices. The cross-attention output $\mathbf{f}_{a2p}$ is computed as:

\begin{equation}
\small
\mathbf{f}_{a2p} = \mathsf{Softmax}\left( \frac{\mathbf{q} \mathbf{k}^\top}{\sqrt{d_k}} \right) \mathbf{v}.
\end{equation}

\noindent This process aligns acoustic features with spatial coordinates, enabling the model to associate audio patterns with geometric contexts. The fused representation $\mathbf{f}_{a2p}$ is then combined with the original audio embeddings via residual connection and layer normalization and gets the $\mathbf{f}_{\text{fuse}}$. Next, we integrate the TDOA embeddings $\mathbf{r}^{emb}_{ij}$ and fusion feature $\mathbf{f}_{\text{fuse}}$ using sparse cross-attention to prioritize salient inter-microphone interactions and get the joint feature $\mathbf{f}_{\text{joint}}$. Different from the traditional cross-attention, the sparse one reduces computational complexity by retaining only the top-$T$ attention weights for each query.

The joint feature $\mathbf{f}_{\text{joint}}$ serves as the input to the masked encoder $\mathcal{E}_{\text{mask}}$. The masked encoder processes $\mathbf{f}_{\text{joint}}$ through a stack of $L$ default transformer blocks. During training, we will randomly mask some tokens, and the corresponding masked inputs are excluded from gradient updates. For a masked token, its feature vector is replaced with a learnable mask embedding. The decoder $\mathcal{D}_{\text{mask}}$ reconstructs predictions for all tokens, including masked ones. Let $\hat{\mathbf{s}}^{emb}$ and $\hat{\mathbf{r}}^{emb}$ denote the reconstructed audio embeddings and spatial coordinates embedding, respectively. The reconstruction process is formalized as:

\begin{equation}
\small
\{\hat{\mathbf{s}}^{emb}, \hat{\mathbf{r}}^{emb}\} = \mathcal{D}_{\text{mask}}\left( \mathcal{E}_{\text{mask}}(\mathbf{f}_{\text{joint}}) \right),
\end{equation}

The final reconstructed audio embeddings $\hat{\mathbf{s}}^{emb}$ will be fed into a simple audio inverse mapper (a two-layer MLP) to generate the reconstructed audio $\hat{\mathbf{s}}$. Meanwhile, the spatial coordinates embedding will undergo further processing using the sparse cross-attention introduced earlier, where it is fused with the audio TDOA embedding $\mathbf{r}^{emb}$. The resulting representation is then input into a position decoder (a two-layer MLP) to predict the sound source location $\{\hat{\mathbf{r}}_k\}_{k=1}^{K}$.

The training loss uses $\mathcal{A}=\mathcal{M}\cup\mathcal{U}$, the set of all microphones with known or unknown positions:

\vspace{-3mm}
\begin{align}
\small
    \mathcal{L}_{\text{total}} = & \ \lambda_{\text{sound}} \sum_{i \in \mathcal{A}} ||\hat{\mathbf{s}}_i - \mathbf{s}_i^{\text{gt}}||_2^2
    + \lambda_{\text{m-loc}} \sum_{i \in \mathcal{A}} ||\hat{\mathbf{r}}_i - \mathbf{r}_i^{\text{gt}}||_2^2 \nonumber \\
    & + \lambda_{\text{s-loc}} \sum_{k=1}^{K} ||\hat{\mathbf{r}}_k - \mathbf{r}_k^{\text{gt}}||_2^2.
\end{align}

%% file: figure/framework.tex
\begin{figure*}[!t]
\centering
\includegraphics[width=1\linewidth]{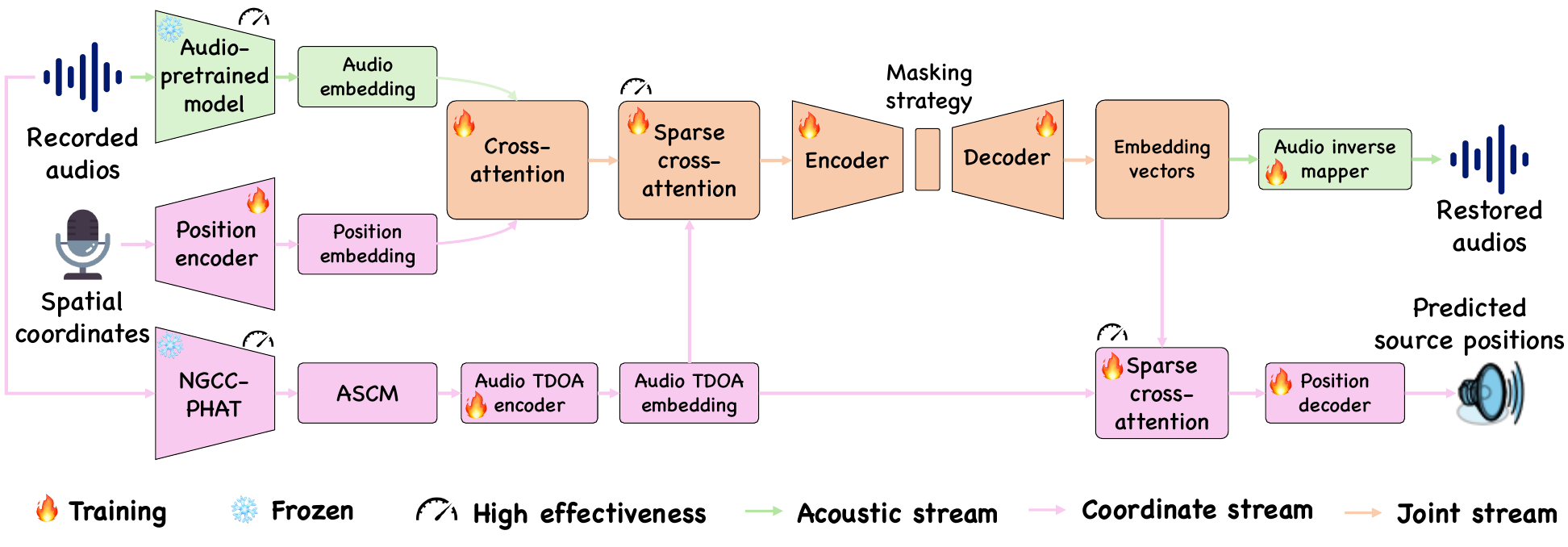} \vspace{-5mm}
\caption{The workflow of the proposed framework. It consists of three streams: \textcolor{LimeGreen}{Acoustic stream} (Section~\ref{AcousticStream}), \textcolor{Thistle}{Coordinate stream} (Section~\ref{CoordinateStream}), and \textcolor{YellowOrange}{Joint stream} (Section~\ref{JointStream}). Within these, we specifically highlight four components that can enhance the efficiency and the parts that need to be trained or frozen during the training process.} 
\vspace{-4mm}
\label{fig:framework} 
\end{figure*}

%% file: 4_experiment.tex
\section{Experiment} \label{sec:Experiment}

\subsection{Experimental Setup}
\textbf{Datasets}: We use two recordings, $\mathsf{music3}$ and $\mathsf{speech3}$, from LuViRA~\cite{yaman2024luvira} audio-only dataset for evaluation and other 3 $\mathsf{music}$ and 3 $\mathsf{speech}$ recordings for training and validation. The sound source is randomly sampled in a room of size $7 \times 8 \times 2$ m. \textbf{Evaluation metrics}: The performance of localization is assessed using Mean Absolute Error (MAE) to quantify overall accuracy. Additionally, we measure the percentage of correctly localized points that fall within a 30 cm threshold to evaluate practical applicability. \textbf{Baselines}: We compare our method with a robust multilateration~\cite{aastrom2021extension} and extended version~\cite{aastrom2021extension,berg2022extending}, existing DI-NN~\cite{grinstein2023dual}, GNN~\cite{grinstein2023graph}, and wav2pos~\cite{berg2024wav2pos}, training them on the same dataset while following their original hyperparameters. \textbf{Deployment}: We train proposed model on one NVIDIA A10 (24 GB). The initial learning rate is 0.001, with a decay of 95\% of every 10 epochs. A batch size of 128 and a training duration of 300 epochs were selected for the proposed model. The model was developed using PyTorch 1.12~\cite{paszke2019pytorch} under Python 3.9\footnote{Thank repo \url{https://github.com/axeber01/wav2pos/}}.

\input{table/main_result}
\input{table/number_mic}
\input{table/ablation_table}
\input{figure/vis}

\subsection{Results}
For the single sound source localization, Table~\ref{mainresult} presents the localization performance of different methods on two recordings using eleven microphones. Our proposed method achieves the best performance across both datasets. Notably, it outperforms wav2pos~\cite{berg2024wav2pos}, the previous best-performing approach, reducing the MAE (cm) to 13.9 for $\mathsf{music3}$ and 21.3 for $\mathsf{speech3}$ while achieving 96.8\% and 88.4\% accuracy, respectively. Our model has 6.8M parameters, with an average training time of 21.5 hours and an inference time of 0.4 seconds based on an A10 GPU. Besides, Table~\ref{mic_ablation} presents the result based on different numbers of microphone configurations. Our approach consistently delivers the lowest MAE. In particular, with seven microphones, it reduces the MAE (cm) to 28.7, which is close to the wav2pos performance using nine microphones. Furthermore, we also conducted extended multiple sound source localization tasks on the $\mathsf{speech3}$ dataset. Using eleven microphones for localization, our method achieved an MAE (cm) of 45.21 for two sound sources and 69.86 for three sound sources.

\subsection{Ablation Study} \label{sec:abtest}
Building on the above foundational experiments, we consider scenes in which some operational microphones have unknown positions. In scene A, both the microphone positions and the sound source are unknown. In scene B, the microphone positions are unknown, but the source audio is known. The results for different microphone counts and methods are shown in Table~\ref{mic_loc}. Our method estimates the unknown microphone positions while localizing the sound source.

Additionally, we visualized the results based on different settings, as shown in Figure~\ref{fig:vis}. Settings 1–3 correspond to the following scenarios: (1) default task, (2) default task with an unknown source signal, and (3) default task with an unknown source signal while also estimating one microphone whose position is unknown.

%% file: table/main_result.tex
\begin{table}[!t]
\centering
\caption{Model localization performance on the LuViRA~\cite{yaman2024luvira} $\mathsf{music3}$ and $\mathsf{speech3}$ recordings using all eleven microphones.}
\resizebox{1\linewidth}{!}{
\fontsize{11}{14}\selectfont
\renewcommand{\arraystretch}{1}{
\begin{tabular}{l|cc|cc}
\toprule
\multicolumn{1}{c|}{\multirow{2}{*}{Method}} & \multicolumn{2}{c|}{$\mathsf{music3}$}& \multicolumn{2}{c}{$\mathsf{speech3}$} \\ \cmidrule{2-5} 
\multicolumn{1}{c|}{} & MAE [cm]$\downarrow$ & acc@30 cm$\uparrow$ & MAE [cm]$\downarrow$ & acc@30 cm$\uparrow$ \\ \midrule
Multilat & 38.8 $\pm$ 2.5 & 72.5 $\pm$ 1.6 & 72.8 $\pm$ 4.4 & 55.7 $\pm$ 2.1 \\
Multilat* & 16.3 $\pm$ 1.6 & 94.7 $\pm$ 0.8 & 34.9 $\pm$ 3.2 & 84.9 $\pm$ 1.6 \\
DI-NN & 26.0 $\pm$ 0.8 & 73.0 $\pm$ 0.2 & 44.7 $\pm$ 1.7 & 45.9 $\pm$ 2.3 \\
GNN & 17.0 $\pm$ 0.7 & 90.7 $\pm$ 1.0 & 31.9 $\pm$ 1.6 & 71.2 $\pm$ 2.0 \\
wav2pos & 14.2 $\pm$ 0.5 & 95.4 $\pm$ 0.7 & 23.6 $\pm$ 1.2 & 81.6 $\pm$ 1.7 \\
\rowcolor{lightgray} \textbf{Ours} & \textbf{13.9 $\pm$ 0.6} & \textbf{96.8 $\pm$ 0.5} & \textbf{21.3 $\pm$ 1.3} & \textbf{88.4 $\pm$ 1.6} \\ \bottomrule
\end{tabular}}} 
\label{mainresult}
\end{table}

%% file: table/number_mic.tex
\begin{table}[!t]
\centering
\caption{Sound source localization MAE [cm] on the $\mathsf{speech3}$ recording using different number of microphones $M$.}
\resizebox{0.92\linewidth}{!}{
\fontsize{11}{14}\selectfont
\renewcommand{\arraystretch}{1}{
\begin{tabular}{l|ccc}
\toprule
Method &$M=5$ & $M=7$ & $M=9$ \\ \midrule
Multilat~\cite{aastrom2021extension} & $244.9 \pm 4.8$ & $133.1 \pm 5.7$ & $94.3 \pm 5.0$ \\
Multilat*~\cite{aastrom2021extension, berg2022extending} & N/A & $105.6 \pm 6.1$ & $56.7 \pm 4.5$ \\
DI-NN~\cite{grinstein2023dual} &$94.9 \pm 2.6$ & $76.1 \pm 2.0$ & $58.5 \pm 1.6$  \\
GNN~\cite{grinstein2023graph}  & $80.5 \pm 2.2$ & $53.1 \pm 1.9$ & $41.1 \pm 1.7$ \\
wav2pos~\cite{berg2024wav2pos} & $66.8 \pm 2.0$ & $38.8 \pm 1.7$ & $28.4 \pm 1.4$ \\
\rowcolor{lightgray} \textbf{Ours} & $\mathbf{42.5 \pm 2.1}$ & $\mathbf{28.7 \pm 1.5}$ & $\mathbf{23.6 \pm 1.2}$ \\ \bottomrule
\end{tabular}}}
\label{mic_ablation}
\end{table}

%% file: table/ablation_table.tex
\begin{table}[!t]
\centering
\caption{Microphone localization MAE [cm] over all unknown microphone locations on the $\mathsf{speech3}$ recordings using different numbers of known microphone locations.}
\resizebox{1\linewidth}{!}{
\fontsize{11}{14}\selectfont
\renewcommand{\arraystretch}{1}{
\begin{tabular}{c|c|ccc} \toprule
Scene & Method & M=7 & M=8 & M=9 \\ \midrule
\multirow{2}{*}{A} & wav2pos & $182.8 \pm 1.7$ & $93.1 \pm 1.9$ & $36.8 \pm 1.9$ \\
 & Ours & $\mathbf{168.9 \pm 1.7}$ & $\mathbf{86.2 \pm 1.9}$ & $\mathbf{31.2 \pm 1.8}$ \\ \bottomrule
\multirow{2}{*}{B} & wav2pos & $181.7 \pm 1.7$ & $90.4 \pm 1.8$ & $34.8 \pm 1.6$ \\
 & Ours & $\mathbf{166.5 \pm 1.8}$ & $\mathbf{83.6 \pm 1.8}$ & $\mathbf{29.5 \pm 1.7}$ \\ \bottomrule
\end{tabular}
}}
\label{mic_loc}
\end{table}

%% file: figure/vis.tex
\begin{figure}[!t]
    \centering
    \subfigure[\small{Setting 1}]{
        \includegraphics[width=0.305\linewidth]{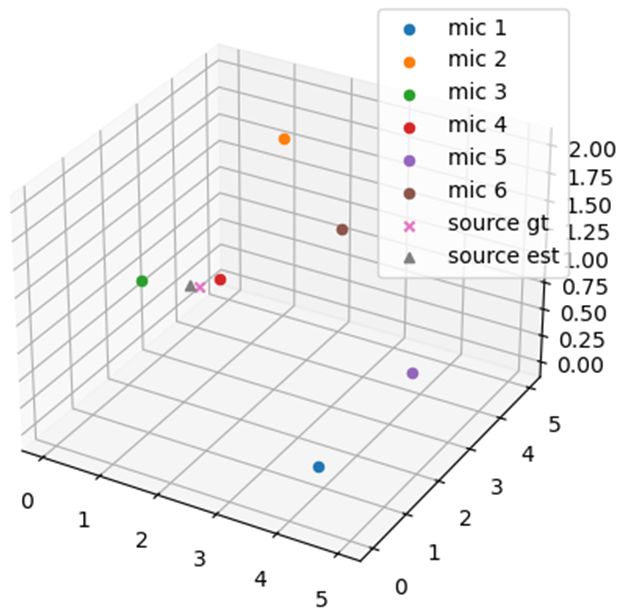}
    }
    \subfigure[\small{Setting 2}]{
        \includegraphics[width=0.305\linewidth]{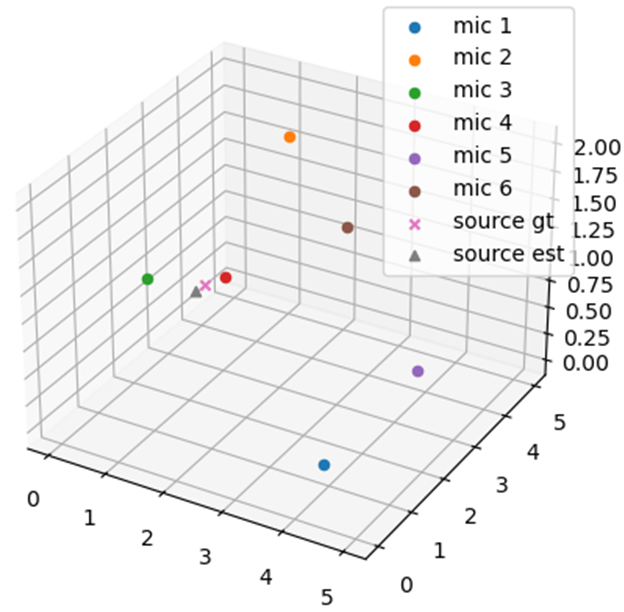}
    }
    \subfigure[\small{Setting 3}]{
        \includegraphics[width=0.305\linewidth]{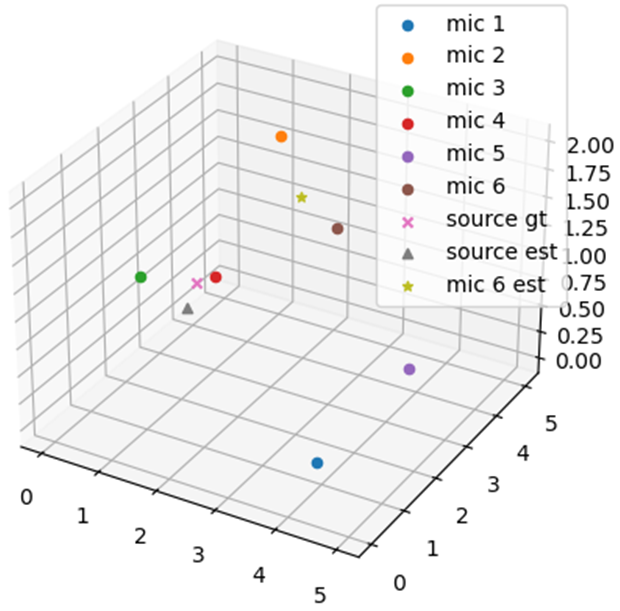}
    }
    \caption{Result visualization under different settings with random initialization of sound source and microphone positions.}
    \label{fig:vis}
\end{figure}

%% file: 5_conclusion.tex
\section{Conclusion} \label{sec:Conclusion}

This paper presented a novel 3D sound source localization framework to address key challenges in efficiency and partially known microphone geometry. By integrating sparse cross-attentions, adaptive signal coherence metrics, and pretrained audio encoders, the proposed method achieves accurate localization with fewer microphones while supporting operational microphones at unknown positions. Its ability to estimate unknown microphone positions and handle overlapping sound events highlights its practicality for real-world deployments.